\documentclass[twocolumn]{aastex631}

\usepackage{amsmath}
\newcommand{\gbar}{g_{\rm bar}}
\newcommand{\gobs}{g_{\rm obs}}
\newcommand{\azero}{a_0}
\newcommand{\Mbar}{M_{\rm bar}}
\newcommand{\Reff}{R_{\rm eff}}
\newcommand{\dmed}{\Delta_{\rm med}}
\newcommand{\dmean}{\Delta_{\rm mean}}
\newcommand{\sigmafloor}{\sigma_{M3}}
\newcommand{\gpred}{g_{\rm pred}}

\shorttitle{The mass--size plane and the RAR identifiability limit}
\shortauthors{Sosna}

\begin{document}

\title{The Mass--Size Plane Does Not Resolve the Identifiability Limit of the
Radial Acceleration Relation}

\author[0009-0004-1266-9729]{Lukas Sosna}
\affiliation{Studio Luki LLC, Millbrae, CA}

\begin{abstract}
We present a computational audit of the identifiability limits for structural corrections to the Radial Acceleration Relation (RAR) using a canonical subset of the SPARC database ($N = 126$). Rather than proposing a new dynamical law, we establish the observational conditions under which such a law would be mathematically recoverable. We resolve three methodological controls that heavily influence RAR interpretations. First, we isolate a $-0.39$ dex residual offset between the eight galaxies in the lowest-quality observational tier ($Q = 3$, mean $-0.433$ dex) and the remaining 118 (mean $-0.040$ dex). Of five predictions that beam smearing must satisfy, four fail: the radial profile plateaus at $-0.331 \pm 0.028$ dex rather than decaying to zero, and the offset does not scale with the number of resolution elements across a curve. Beam smearing is disfavoured as the primary driver, though the offset remains a data-quality signature rather than a physical one --- $Q = 3$ in SPARC flags major asymmetries and strong non-circular motions, conditions under which a rotation curve does not trace the equilibrium potential. Second, we decouple the architectural limits of the dataset into three independently measured quantities --- the no-model point scatter ($\sigma_{M0} = 0.1860$ dex), the residual floor after free per-galaxy intercepts ($\sigma_{M3} = 0.1058$ dex), and the propagated analytic error floor --- together with the absorbable budget $\sqrt{\sigma_{M0}^2 - \sigma_{M3}^2} = 0.1530$ dex derived from the first two. Finally, we provide an 8-cell protocol grid to reconcile Leave-One-Out (LOO) Mean Squared Prediction Error (MSPE) ratios. We demonstrate that reported structural--dynamical couplings must be evaluated with strict adherence to residual definitions (median vs. mean, signed vs. absolute) and baseline denominators so that labeling artifacts are not mistaken for physical effects.
\end{abstract}


\section{Introduction}
\label{sec:intro}

The radial acceleration relation \citep{mcgaugh2016,lelli2017} ties the observed
centripetal acceleration $\gobs$ of a galaxy's rotation curve to the
acceleration $\gbar$ implied by its baryons alone, with a scatter small enough
that the relation has been read as evidence for a modified dynamical law. Whether
the residual scatter about the relation carries structural information --- a
dependence on galaxy size, surface density, or compactness at fixed baryonic
mass --- is the question that separates a law with an intrinsically small scatter
from a law plus an unmodelled structural term.

The companion audit \citep{sosna2026} posed that question as an identifiability
problem rather than a detection problem. It asked, of each candidate
one-parameter structural coordinate, three questions in sequence: does it
out-predict a mass-only baseline out of sample; does it clear the scatter floor
that remains once every galaxy is granted its own free intercept; and does it
survive stratification by data quality after mass is controlled. Compactness
$\lambda = G\Mbar/\Reff c^2$ --- the strongest candidate of the family --- failed
the first gate and was outscored by the SPARC quality flag $Q$. The audit's
conclusion was deliberately negative and deliberately narrow: a one-parameter
structural correction of this form is not identifiable in this sample.

A narrow negative result is open to a natural objection. Perhaps the coordinate
family was too small. If the residual depends on mass and size through a genuine
two-dimensional surface, then every one-parameter projection of that surface is
a lossy summary, and failing to identify any single projection says nothing
about the surface itself. This paper tests that objection directly.

We keep the audit's sample, protocol, residual definitions, and scoring rules
fixed and change only the dimension of the correction. Section~\ref{sec:data}
states the frozen protocol and the reporting discipline we adopt for
correlations. Section~\ref{sec:manifold} is the main test. Section~\ref{sec:qflag}
asks what the quality flag is actually measuring, since it is the coordinate that
beats every physical one. Sections~\ref{sec:floors} and \ref{sec:q3} close two
bookkeeping questions that a referee would otherwise raise: whether the several
scatter floors quoted in this literature are the same quantity, and whether the
lowest-quality galaxies are a resolution artefact. Section~\ref{sec:anchors}
reports two external anchors that do not settle the question, and
Section~\ref{sec:settle} names the measurement that would.

\section{Data and frozen protocol}
\label{sec:data}

\subsection{Sample}

We use the SPARC database \citep{lelli2016} with the sample definition frozen by
the companion audit: inclination $30^\circ < i < 80^\circ$, quality flag
$Q \leq 3$, and at least five rotation-curve points per galaxy. This yields
$N = 126$ galaxies and $2709$ rotation-curve points, split at the median of
$\log \gbar = -10.926$ into $63$ low-acceleration and $63$ high-acceleration
galaxies. The quality flag distribution is $Q=1$: $74$, $Q=2$: $44$, $Q=3$: $8$.
All $31$ numerical anchors of the companion audit regenerate from the raw tables
through the shared sample builder used here
(\texttt{results/reproducibility\_receipt.json}).

The RAR residual at a rotation-curve point is
\begin{equation}
\Delta = \log_{10} \gobs - \log_{10} \gpred,
\qquad
\gpred = \frac{\gbar}{1 - e^{-\sqrt{\gbar/\azero}}},
\label{eq:resid}
\end{equation}
with $\azero = 1.2 \times 10^{-10}\ {\rm m\ s^{-2}}$ held fixed at the
literature value \citep{mcgaugh2016}. We aggregate to one number per galaxy two
ways --- the median $\dmed$ and the mean $\dmean$ of that galaxy's points --- and
report both throughout, because they are not interchangeable and the difference
between them has been a source of confusion in this problem
(Section~\ref{sec:protocol-triple}).

\subsection{The protocol triple}
\label{sec:protocol-triple}

\begin{figure*}
\centering
\includegraphics[width=\textwidth]{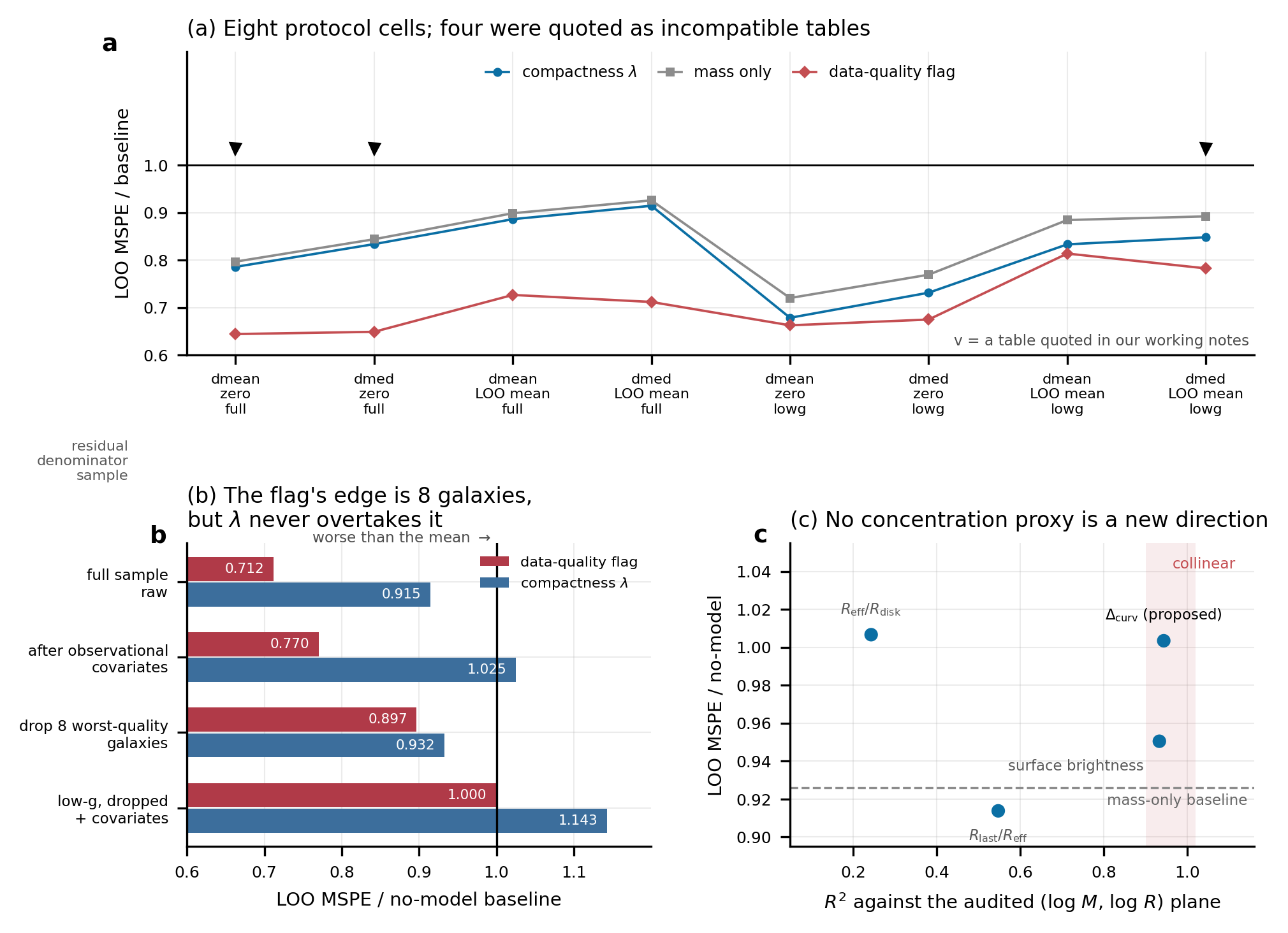}
\caption{The quality flag against the physical coordinates. Panel (a) shows that
tables which appear mutually incompatible are one pipeline evaluated under eight protocol cells, four of which correspond to tables quoted in our working notes; panel (b) that the flag's edge narrows but does not vanish when the
eight lowest-quality galaxies are dropped; panel (c) that no concentration proxy
supplies a direction independent of the audited plane.
\label{fig:qaqc}}
\end{figure*}

Correlations in this problem are unusually sensitive to protocol choices that
are easy to leave unstated. Over the course of this work we found that a single
pair of quoted correlation coefficients could not be reproduced until the
residual definition, the statistic, and the sample were all specified together,
and that two numbers quoted as a pair in fact required different quality cuts
(Section~\ref{sec:anchors-recon}). We therefore adopt a reporting rule for this
paper and recommend it generally: no correlation is reportable without its
\emph{protocol triple} --- the residual definition, the statistic including any
controlled variables, and the sample including every cut. Every correlation
below carries its triple, and the machine-readable results files record it per
value. The grid has eight cells: two residual definitions ($\dmed$, $\dmean$) $\times$ two denominators (literal-zero predictor, leave-one-out grand mean) $\times$ two samples (full, low $g$). Four of the eight correspond to tables that circulated in our working notes as apparently incompatible; all four reproduce to within $0.0004$ once their cell is stated (Figure~\ref{fig:qaqc}a).

\section{The two-dimensional mass--size surface}
\label{sec:manifold}

\subsection{Interpretation of the nuisance floor}
\label{sec:floor-logic}

The per-galaxy nuisance floor $\sigmafloor = 0.1058$ dex is the point-level
residual scatter that remains after granting every galaxy its own free
intercept. We state the logic explicitly, since the floor is easily misread as a benchmark for models to clear. A
free per-galaxy offset is the saturated galaxy-level model. Any function of
$(\log \Mbar, \log \Reff)$ is a constrained special case of it, so the
point-level scatter of any such function is bounded \emph{below} by
$\sigmafloor$ as a matter of algebra rather than physics. An apparent violation of this bound would therefore indicate a computational error rather than a physical result.

The decidable question is what fraction of the \emph{absorbable} per-galaxy
budget (the scatter the free intercepts remove) a declared structural
function explains. That is the quantity we report.

\subsection{Size dependence at fixed mass}

Table~\ref{tab:partials} gives the partial correlation of $\log \Reff$ with the
residual controlling for $\log \Mbar$.

\begin{deluxetable}{lrrl}
\tablecaption{Partial correlation of $\log \Reff$ with $\dmed$, controlling
$\log \Mbar$. Sample: frozen protocol, $N$ as stated.\label{tab:partials}}
\tablewidth{0pt}
\tablehead{\colhead{Regime} & \colhead{$N$} & \colhead{partial $r$} &
\colhead{$p$}}
\startdata
full & 126 & $-0.113$ & 0.210 \\
low $g$ & 63 & $\mathbf{-0.287}$ & $\mathbf{0.024}$ \\
high $g$ & 63 & $+0.023$ & 0.860 \\
\enddata
\end{deluxetable}

Size adds information at fixed mass in the low-acceleration regime and nowhere else. Correlations between RAR residuals and galaxy properties have been examined before: \citet{li2018} fit the relation to individual SPARC galaxies; \citet{stone2019} recovered an intrinsic scatter of $0.11 \pm 0.02$ dex for the stellar RAR through a Monte Carlo error model, closely bracketing the floor we measure here; \citet{desmond2023} marginalised over distance, inclination, luminosity, and mass-to-light ratio to obtain an underlying scatter of $0.034 \pm 0.002$ dex; and \citet{stiskalek2023} tested residual dependence on a range of galaxy variables and concluded that the RAR is the only fundamental correlation of late-type galaxy dynamics. Our $\sigmafloor$ is not directly comparable to Desmond's figure --- a saturated per-galaxy intercept and a marginalised hierarchical fit are different estimators, and the intercept absorbs coherent structure the marginalisation models explicitly --- but the ordering is consistent. None of these reports a size dependence at fixed mass confined to the low-acceleration half of the sample, and we are not aware of a prior treatment framed as identifiability rather than detection. The finding is quality-sensitive (Section~\ref{sec:quality-strat}); we present it with that caveat attached rather than in a robustness appendix.

The two axes are strongly collinear ($r = 0.766$ between $\log \Mbar$ and
$\log \Reff$ at low $g$, $58.6\%$ shared variance, VIF $= 2.42$), which is the
proximate reason the two-dimensional fit buys little over one dimension.

\subsection{Out-of-sample scoring}
\label{sec:loo}

Table~\ref{tab:loo} scores each model by leave-one-out mean squared prediction
error, normalised to a no-model baseline, in the low-acceleration regime where
the effect lives.

\begin{deluxetable}{lr}
\tablecaption{Leave-one-out MSPE relative to the no-model baseline, low $g$
($N = 63$), residual $\dmed$. Lower is better; ratios below unity beat the
baseline.\label{tab:loo}}
\tablewidth{0pt}
\tablehead{\colhead{Model} & \colhead{MSPE / no-model}}
\startdata
size only & 1.014 \\
surface density $\Sigma_{\rm eff}$ & 0.909 \\
mass only (companion baseline) & 0.892 \\
2D mass--size surface & 0.869 \\
compactness $\lambda$ (1 parameter) & 0.848 \\
data-quality flag $Q$ alone & \textbf{0.783} \\
\enddata
\end{deluxetable}

Three conclusions follow from Table~\ref{tab:loo}.

First, the two-dimensional surface beats mass-only, but by $2.6\%$
(ratio $0.974$ against mass-only) with permutation $p = 0.068$ over $600$
permutations --- not significance at any pre-registered threshold.

Second, despite the additional free parameter, the surface performs worse than a one-parameter declared coordinate. Compactness $\lambda$ scores $0.848$ against
the surface's $0.869$. Adding a dimension did not help; the surface's fitted
direction simply rediscovers $\lambda$ (Section~\ref{sec:direction}).

Third, both lose to the data-quality flag at $0.783$. A three-level integer
recording how much an observer trusted a rotation curve out-predicts every
physical coordinate tested. Section~\ref{sec:qflag} examines this result in detail.

Outside the low-acceleration regime the surface is worse than mass-only: ratios
$1.003$ (full, $p = 0.196$) and $1.028$ (high $g$, $p = 0.747$).

A note on model classes. A one-dimensional projection with a \emph{fitted} slope
spans the same function class as the two-dimensional linear model, so scoring
``2D versus best-fitted-1D'' compares a model against itself. All comparisons
above are therefore against \emph{declared} directions --- mass, size, $\lambda$,
$\Sigma_{\rm eff}$ --- which is the only form of the test that carries
information.

\subsection{The fitted direction is compactness}
\label{sec:direction}

The best-fit direction in the $(\log \Mbar, \log \Reff)$ plane at low $g$ is
$\theta = -48.9^\circ$, with bootstrap confidence intervals
CI68 $[-54.5^\circ, -39.4^\circ]$ and CI95 $[-59.4^\circ, -10.9^\circ]$.
Compactness $\lambda = G\Mbar/\Reff c^2$ sits at exactly $-45^\circ$, inside
CI68. The fitted direction is statistically indistinguishable from the
coordinate the companion audit already tested and could not attribute. In the
full sample the direction is poorly constrained (CI95 spans $121^\circ$); at
high $g$ it is unconstrained over essentially the whole circle.

The fitted direction therefore recovers the coordinate already tested rather than identifying a new one.

\subsection{The nuisance budget}

\begin{deluxetable*}{lcccc}
\tablecaption{Point-level scatter and the fraction of the absorbable per-galaxy
budget explained by the 2D surface. Residual $\dmed$, frozen
protocol.\label{tab:budget}}
\tablewidth{0pt}
\tablehead{\colhead{Regime} & \colhead{$\sigma$ (no model)} &
\colhead{$\sigma$ (2D)} & \colhead{$\sigmafloor$} &
\colhead{budget explained}}
\startdata
full & 0.186 & 0.177 & 0.1058 & 10.7\% \\
low $g$ & 0.230 & 0.220 & 0.1271 & 10.1\% \\
high $g$ & 0.148 & 0.147 & 0.0921 & 1.5\% \\
\enddata
\tablecomments{$\sigmafloor$ is evaluated within each regime; the full-sample
value is the constant of Table~\ref{tab:floors}. Budget explained is
$(\sigma_{\rm no\ model} - \sigma_{\rm 2D})/(\sigma_{\rm no\ model} -
\sigmafloor)$, all three evaluated within the regime.}
\end{deluxetable*}

\begin{figure*}
\centering
\includegraphics[width=\textwidth]{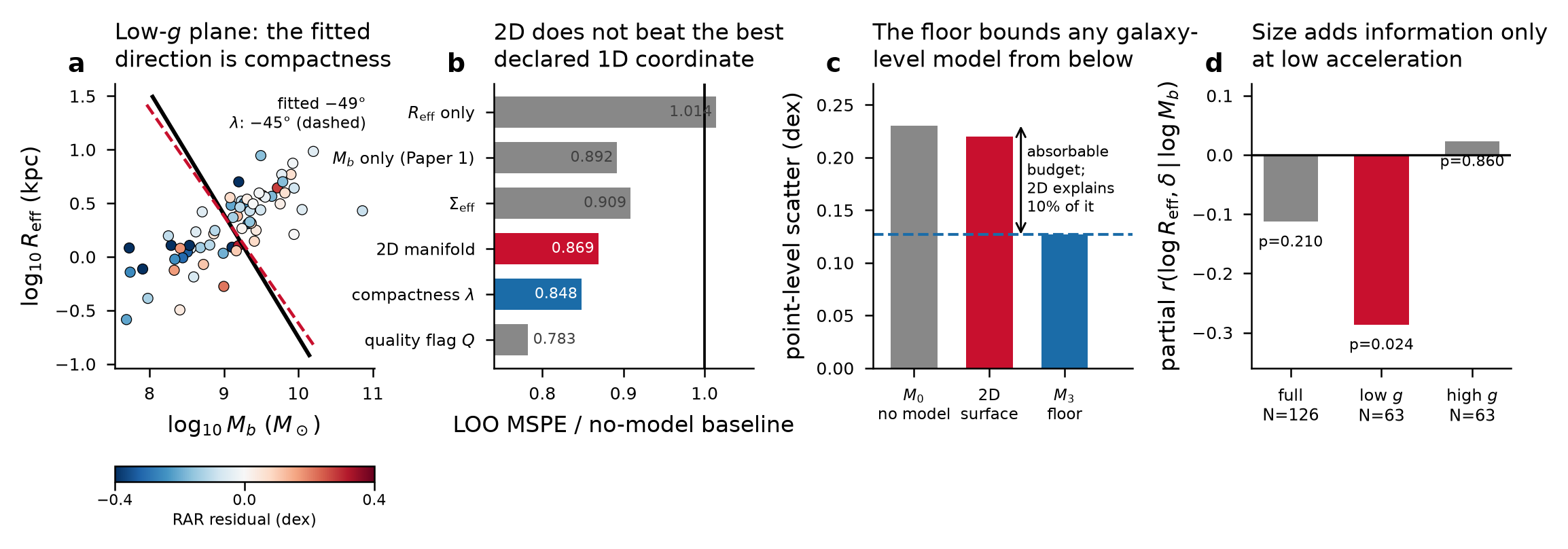}
\caption{The two-dimensional mass--size test. The fitted direction at low
acceleration is statistically indistinguishable from compactness, and the
surface explains about a tenth of the absorbable per-galaxy nuisance budget.
\label{fig:manifold}}
\end{figure*}

Table~\ref{tab:budget} gives the point-level scatter by regime, and
Figure~\ref{fig:manifold} summarises the test as a whole. The surface explains
about $10\%$ of the absorbable per-galaxy budget at full sample and low $g$, and
essentially nothing at high $g$. Roughly $90\%$ of the
per-galaxy freedom remains attributable to distance, inclination, and
mass-to-light nuisance rather than to structure , a decomposition we make quantitative in Section~\ref{sec:floors}.

\subsection{Quality stratification}
\label{sec:quality-strat}

The low-$g$ result does not survive stratification by quality flag. Splitting the
low-acceleration sample:

\begin{itemize}
\item $Q = 1$ ($N = 31$, best data): ratio $0.947$, $p = 0.062$
\item $Q = 2$ ($N = 27$): ratio $\mathbf{1.154}$, $p = 0.902$ --- the surface is
\emph{worse} than mass-only
\end{itemize}

The effect lives entirely in the highest-quality tier and inverts in the next
one. Combined with $Q$ outscoring every physical coordinate, this is the hazard
the companion audit identified, now reproduced with an extra parameter: the
residual structure tracks data quality at least as well as it tracks physics.

The result is robust to the residual definition without becoming significant:
ratio $0.974$ ($p = 0.068$) for $\dmed$, $0.964$ ($p = 0.065$) for $\dmean$, and
$0.975$ ($p = 0.078$) under an asymmetric-drift correction with
$\sigma = 8\ {\rm km\ s^{-1}}$.

\section{The information content of the quality flag}
\label{sec:qflag}

That a three-level integer out-predicts every physical coordinate requires explanation.
Two are available: $Q$ is a proxy for measurement error, in which case it
belongs in the noise model and not in the physics; or $Q$ carries information
about galaxy structure that the structural coordinates fail to capture, in which
case the identifiability problem is worse than a coordinate choice. We ran five
tests to separate these.

\emph{Permuted labels.} Randomly relabelling galaxies with the same
three-way group sizes gives mean ratio $1.017$ with 5th percentile $0.987$; the
real labelling's $0.712$ (full sample) lies outside the null,
$p < 1/600$ over the permutations run. The flag's edge is not an artefact of
fitting three group means.

\emph{Mass control.} Residualising on $\log \Mbar$ first, $Q$ retains a ratio of
$0.779$ while $\lambda$ (1.022) and the 2D surface (1.027) both become worse than
the baseline. Adding observational covariates leaves $Q$ at $0.770$.

\emph{Matched comparison.} Scored as three binned group means --- the same
functional form $Q$ is granted --- the physical coordinates give $0.959$
($\log\lambda$), $0.974$ ($\log \Mbar$), and $0.981$ ($\log \Reff$) against $Q$'s
$0.712$. The advantage is not a functional-form artefact.

\emph{What $Q$ correlates with.} In the full sample, Spearman
$\rho(Q, \log \Reff) = -0.406$ ($p = 2.5\times10^{-6}$),
$\rho(Q, \log \Mbar) = -0.367$ ($p = 2.4\times10^{-5}$),
$\rho(Q, \log\lambda) = -0.290$ ($p = 9.8\times10^{-4}$),
$\rho(Q, N_{\rm pts}) = -0.308$ ($p = 4.6\times10^{-4}$), and
$\rho(Q, \sigma_i) = +0.138$ ($p = 0.125$, where $\sigma_i$ is the inclination uncertainty). Low-quality galaxies are
preferentially small, low-mass, and sparsely sampled --- so $Q$ is entangled with
structure, not orthogonal to it.

\emph{The measurement block.} This test discriminates between the two hypotheses. If $Q$ were merely an
error proxy, a model built from the measurement uncertainties themselves should
approach its performance. It does not: the measurement block scores $0.925$
against $Q$'s $0.712$, and combining them ($0.724$) does not improve on $Q$
alone. Whatever $Q$ encodes is not captured by the published uncertainties.

$Q$ is therefore neither a pure error proxy nor a cleanly structural quantity: it is an observer's judgement correlated with both, and its predictive advantage over every declared physical coordinate is the strongest single constraint this dataset places on the residual field.

\section{Three distinct scatter floors}
\label{sec:floors}

\begin{figure*}
\centering
\includegraphics[width=\textwidth]{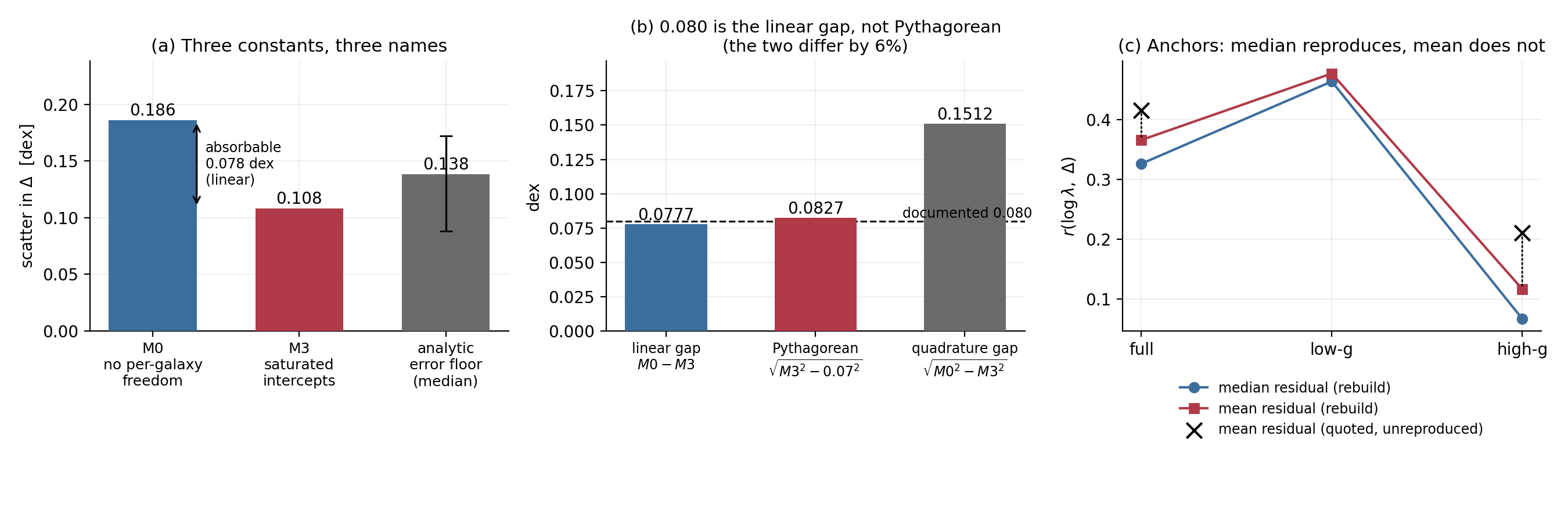}
\caption{Left: split by channel, the propagated error model accounts for both
the post-intercept floor and the absorbable budget. Right: the resolution of a
disputed pair of correlation coefficients, which differ from the frozen-protocol
values because they were computed on the absolute residual with the sign
dropped.
\label{fig:floors}}
\end{figure*}

The literature on this problem, including our own working notes, has quoted
several scatter floors under a single name. They are three distinct floors ---
$\sigma_{M0}$, $\sigmafloor$, and the propagated analytic floor, the last
resolving into two channels --- plus one quantity derived from the first two,
the absorbable budget. Conflating them produces a spurious inference. We fix
them by name in Table~\ref{tab:floors}, and Figure~\ref{fig:floors} shows the
channel split against the measured floors.

\begin{deluxetable}{llc}
\tablecaption{The floor constants and the propagated error budget.\label{tab:floors}}
\tabletypesize{\footnotesize}
\tablewidth{0pt}
\tablehead{\colhead{Constant} & \colhead{Definition} & \colhead{Value}}
\startdata
$\sigmafloor$ & scatter after free per-galaxy intercepts & 0.1058 \\
$\sigma_{M0}$ & scatter with no galaxy-level model & 0.1860 \\
absorbable & $\sqrt{\sigma_{M0}^2 - \sigmafloor^2}$ & 0.1530 \\
analytic, within-galaxy & propagated velocity errors & 0.1000 \\
analytic, coherent & propagated $D$, $i$, $M/L$ errors & 0.1306 \\
\enddata
\tablecomments{All values in dex.}
\end{deluxetable}

The inference that matters concerns whether the propagated measurement-error
budget accounts for these. It does, but only once the budget is split by
\emph{channel}. Distance and inclination errors shift every point within a
galaxy coherently (they are exactly what a free per-galaxy intercept absorbs) while velocity errors vary point to point within a galaxy. A single-number
error budget containing both cannot be compared against $\sigmafloor$, which is
measured \emph{after} galaxy means are removed, without double-counting the
coherent terms.

Split by channel:
\begin{itemize}
\item within-galaxy channel: predicted $0.100$ dex against observed
$\sigmafloor = 0.1058$ dex, ratio $1.058$;
\item coherent channel (distance, inclination, mass-to-light): predicted
$0.1306$ dex against the absorbable budget in quadrature $0.153$ dex, ratio
$1.171$.
\end{itemize}

Both floors are accounted for by the error model to within about $6\%$ and
$17\%$ respectively. In particular there is no residual requirement for
unexplained point-level structure inside galaxies; an earlier reading to that
effect rested on comparing a channel-conflated budget against $\sigmafloor$.
That the independent-error sum slightly over-predicts the coherent channel is
expected, since velocity errors along a rotation curve are correlated and an
independent sum over-counts.

We also tested whether the near-equality
$\sqrt{\sigma_{M3}^2 - (0.07)^2} \approx 0.080$, which had been used to argue
that two of the three constants are not independent, is
construction or coincidence. The near-equality does not survive: the $0.07$ dex input does not reproduce under the frozen protocol, so the relation it appeared to establish has no referent. The constants are independently defined.

\subsection{Resolution of a disputed pair of correlation coefficients}
\label{sec:anchors-recon}

Two correlation coefficients for the $\log\lambda$--$\dmean$ coupling circulated
in our working notes as $0.416$ (full) and $0.211$ (high $g$), against
independently recomputed values of $0.366$ and $0.117$. Rather than selecting between them, we enumerated the cross-product of every protocol axis that fragments these
correlations --- residual definition, sample regime, statistic including a
partial correlation controlling for mass, inclination cut, quality cut, and
minimum points per galaxy --- recomputing $\lambda$ inside each configuration.

The frozen protocol gives $0.366$ (full), $0.477$ (low $g$), $0.117$ (high $g$)
for $\dmean$, reproducing the independent recomputation exactly. The disputed
$0.416$ is $r(\log\lambda, |\dmean|) = -0.4168$ --- the \emph{absolute} residual,
reported with its sign dropped. The distinction is material: it changes the question from whether $\lambda$ predicts the direction of a galaxy's offset to
whether it predicts only its magnitude, and the absolute-residual correlation
carries the opposite sign, so quoting it as $+0.416$ inverts the relationship.
The disputed $0.211$ requires the absolute residual \emph{and} $Q \leq 2$ rather
than the frozen $Q \leq 3$, and is reproduced by no configuration under the
frozen protocol. The two values therefore require different protocols and are
not a coherent pair. The frozen-protocol values are retained.

\section{The eight lowest-quality galaxies}
\label{sec:q3}

\begin{figure*}
\centering
\includegraphics[width=\textwidth]{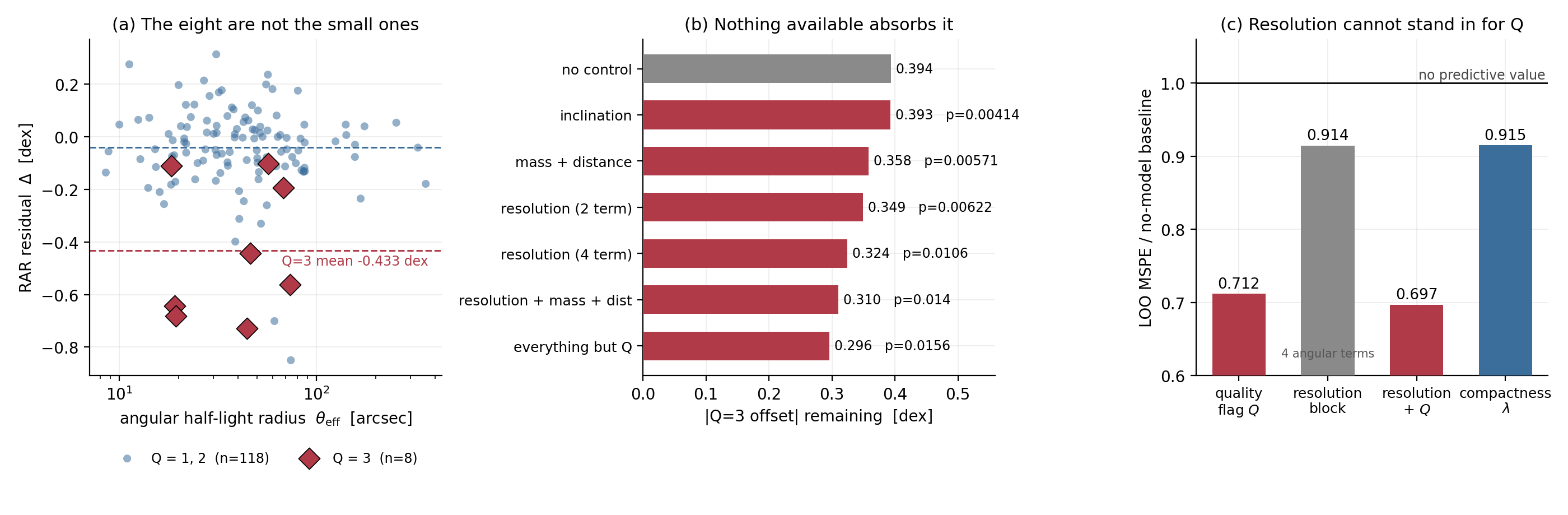}
\caption{Left: the $Q=3$ residual profile in resolution elements plateaus rather
than decaying to the zero line beam smearing would require. Right: dropping the
eight galaxies costs a third of the $Q$--residual correlation and nothing of the
$\lambda$ coupling.
\label{fig:q3}}
\end{figure*}

Eight galaxies carry $Q = 3$ and a mean per-galaxy median residual of $-0.433$
dex, against $-0.040$ dex for the other $118$ --- four times $\sigmafloor$. The
offset is negative, which is the direction beam smearing predicts, since a
finite beam flattens the inner rise of a rotation curve and so underestimates
$\gobs$ where the curve is steep. Sign alone does not discriminate. SPARC
publishes no beam sizes, so we use the median radial sampling step $\delta R$ as
a proxy for the resolution element and $n_{\rm beam} = R_{\rm last}/\delta R$ as
the number of elements across a curve, and test five predictions that beam
smearing must satisfy (Table~\ref{tab:q3}; Figure~\ref{fig:q3}).

\begin{deluxetable*}{ll}
\tablecaption{Beam-smearing predictions for the $Q=3$ offset. Frozen protocol,
residual $\Delta$ at the point level except where noted.\label{tab:q3}}
\tablewidth{0pt}
\tablehead{\colhead{Prediction} & \colhead{Result}}
\startdata
offset scales with $n_{\rm beam}$ within $Q = 3$ & $r = +0.227$, $p = 0.59$: no \\
radial profile decays to zero & plateaus at $-0.331 \pm 0.028$ dex: no \\
offset shrinks when inner points cut & $-0.4332 \to -0.4213$: no \\
matched controls show it too & $-0.110$ vs $-0.433$, $p = 0.009$: no \\
positive radial slope & $+0.152 \pm 0.065$, $p = 0.021$: yes \\
\enddata
\tablecomments{The outer plateau differs from zero at $z = -11.8$. Beam smearing
requires decay to zero, not to a finite plateau.}
\end{deluxetable*}

One of five holds. The profile does have a mild inner gradient --- about $23\%$
of the offset --- but $77\%$ persists at large radius, and beam smearing requires
decay to zero, not to $-0.33$ dex. The flag survives control for acceleration,
mass, and resolution jointly ($r = -0.469$, $p = 3\times10^{-8}$), and within the
low-acceleration half alone the $Q=3$ galaxies sit at $-0.437$ against $-0.071$
for $Q \leq 2$. UGC04305 provides a counterexample: $22$ resolution elements
across its curve and still $-0.562$ dex.

Beam smearing is therefore disfavoured as the primary driver. This does not make
the offset physical. $Q = 3$ in SPARC flags major asymmetries, strong
non-circular motions, and stellar--gas offsets --- conditions under which a
rotation curve does not trace the equilibrium potential at all. That is still a
data-quality artefact, but a different one, and it is not repaired by a
resolution cut.

\subsection{The leverage of the $Q = 3$ galaxies}

The eight galaxies raise a natural concern: that they carry enough of the coupling amplitude to threaten the central claim. The concern correctly identifies a dependence but misattributes it. Dropping them
costs $35\%$ of $|r(Q, \Delta)|$ --- which is close to tautological, since $Q=3$
is the extreme of the $Q$ scale and removing it truncates the predictor --- and
costs nothing of the $\lambda$ coupling: $r(\log\lambda, \dmean)$ moves from
$0.3660$ to $0.3681$ and $r(\log\lambda, \dmed)$ from $0.3263$ to $0.3196$. The
quoted values $-0.351$ raw and $-0.280$ controlling for mass, $p = 0.0022$,
reproduce exactly as $r(Q, \dmed)$ on the $Q \leq 2$ subsample, which confirms
the referent. The caveat applies to the $Q$--residual correlation; it does not transfer to the $\lambda$ coupling.

\section{Two inconclusive external anchors}
\label{sec:anchors}

\subsection{Resolved asymmetric drift in gas-dominated dwarfs}

Nine LITTLE THINGS galaxies with resolved asymmetric-drift corrections
\citep{oh2015} are available in the companion archive. Training both models on
the other $117$ galaxies and predicting these nine out of sample gives RMSE
$0.196$ dex for the 2D surface, $0.194$ dex for mass-only, and $0.186$ dex for a
constant. The constant wins, all three are indistinguishable at $N = 9$, and all
exceed the floor. Excluding the highest-leverage system leaves the ordering
unchanged. We record this anchor as underpowered: the standard error on an RMSE difference at $N=9$ with $0.19$ dex
scatter is far larger than the $0.002$ dex separation observed. It neither supports nor refutes the surface.

\subsection{Dwarf spheroidals}

The Local Group dwarf spheroidals \citep{mcconnachie2012} extend $1.23$ dex below
the SPARC low-$g$ median in acceleration, which makes them the natural lever arm.
Applying the \citet{wolf2010} mass estimator to $46$ systems, the
mass--degeneracy result replicates exactly --- MSPE ratio $1.000$, correlations
identical to six decimal places on a different dataset with a different mass
estimator. The agreement is exact because the degeneracy is algebraic rather than empirical: any monotone reparametrisation of mass carries the same information as mass. However, the sample cannot test the structural question: $89\%$ of it is
external-field dominated, and the residuals are positive
(mean $+0.504$, median $+0.373$, sd $0.483$ dex), consistent with the
external-field effect rather than with a structural term. We report the replication but do not use this sample as an anchor.

\section{A measurement that would break the degeneracy}
\label{sec:settle}

The measurement that breaks this degeneracy is resolved gas velocity dispersion
$\sigma_{\rm gas}(R)$ in gas-dominated dwarfs. The reason is specific: the
per-galaxy free intercept currently absorbs both the pressure-support correction
and any genuine structural offset, and no re-coordinatisation of SPARC can
separate them because both enter as a single galaxy-level constant. A resolved
dispersion profile constrains the pressure-support term directly, leaving the
structural offset identifiable as what remains.

We note the power limitation explicitly. The isolated dwarf spheroidal
spectroscopic test that would provide an independent regime has
$N \approx 5$--$10$ usable systems, which is not enough to detect an effect of
the size reported here. We therefore identify it as a candidate for pre-registration rather than as a present constraint.

\section{Conclusions}
\label{sec:conclusions}

Promoting the structural correction from one parameter to two does not resolve
the identifiability limit established in the companion audit.

\begin{enumerate}
\item The two-dimensional mass--size surface improves out-of-sample prediction
over mass-only by $2.6\%$, in the low-acceleration regime only, at permutation
$p = 0.068$; it is worse than mass-only in the full and high-acceleration
samples.
\item It is beaten by the one-parameter compactness coordinate $\lambda$
($0.869$ versus $0.848$) and by a wider margin by the data-quality flag
($0.783$).
\item Its fitted direction, $\theta = -48.9^\circ$ with CI68
$[-54.5^\circ, -39.4^\circ]$, is statistically indistinguishable from
$\lambda$'s $-45^\circ$. The added dimension recovers the previously tested coordinate.
\item It explains $\sim 10\%$ of the absorbable per-galaxy nuisance budget and
inverts sign between quality tiers.
\item Size does carry information at fixed mass at low acceleration alone
(partial $r = -0.287$, $p = 0.024$). This is the paper's one positive result and
it is bounded and quality-sensitive.
\item The quality flag's advantage is not reducible to a measurement-error
proxy: a model built from the published uncertainties scores $0.925$ against the
flag's $0.712$.
\end{enumerate}

The identifiability boundary is dimensional, not coordinate-specific. It is not
lifted by adding a second structural parameter, as the scoring shows: the residual field's strongest predictor in this dataset remains an
observer's quality judgement.

\begin{acknowledgments}
Portions of the manuscript text were drafted and edited with the assistance of
large language models (Anthropic Claude), which were also used to develop and
review analysis code and to cross-check numerical results. All quantitative
results derive from the archived analysis pipeline and regenerate from the raw
SPARC tables via \texttt{python code/verify\_all.py}, independently of any
language model. The author reviewed all content and takes full intellectual
responsibility for the analysis, results, and conclusions presented here.
\end{acknowledgments}

\software{numpy \citep{harris2020}, scipy \citep{virtanen2020},
matplotlib \citep{hunter2007}}

\section*{Data availability}

The canonical dataset ($N=126$), analysis code, and QA/QC pipeline (including nuisance floor bounds and data-quality leverage controls) supporting this work are permanently archived and computationally reproducible at \url{https://doi.org/10.5281/zenodo.21959872}. Every number quoted above regenerates from the raw SPARC tables via \texttt{python code/verify\_all.py}.

\end{document}